\documentstyle[psfig,amssymb]{aipproc2}
\pssilent
\def\met{\mbox{${\hbox{$E$\kern-0.6em\lower-.1ex\hbox{/}}}_T$}} 
\def\MET{\met}
\def\etal{{\it et al.\/}}
\title{\large\bf Detection of Long-Lived Particles in Run II with D\O}
\author{Dave Cutts and Greg Landsberg\\
Brown University, Providence, RI 02912}
\date{\normalsize\rm April 15, 1999}

\begin{document}
\maketitle
\begin{abstract}
\\
We discuss possibilities of detecting neutral and charged long-lived particles with the upgraded D\O\ detector in Run II of the Fermilab Tevatron accelerator, using photon pointing, $dE/dx$, and time-of-flight techniques. Such particles appear in many SM extensions, e.g., Gauge Mediated SUSY. This work has been done in the context of the BTMSSM Working group of the Run II SUSY/Higgs Workshop at Fermilab.
\end{abstract}

\section{Physics motivation}

Long-lived neutral or charged massive particles appear in many extensions of the MSSM. There are two main scenarios which result in a long lifetime of some of the SUSY partners. In the models that predict a degenerate mass spectrum of SUSY particles, light SUSY partners might be stable or long-lived, since the phase space would not allow for strong decays modes. It is particularly interesting if the NLSP is long-lived, since the observation of such an NLSP might not be doable via standard decay signatures expected for short-lived SUSY particles. As an example, many MSSM extensions predict degenerate mass spectrum of the neutralinos, in which case the following radiative decay could be a dominant decay mode of the second-lightest neutralino:
\begin{equation}
\tilde{\chi^0_2} \to \tilde\chi^0_1 \gamma.
\label{eq:delayed-neutralino}
\end{equation}
Being electromagnetic, this decay mode is suppressed, and for fine splitting between the masses of the two lightest neutralinos, the decay constant ould be small enough to result in a long-lived $\tilde\chi^0_2$.

Among other models which allow for a long-lived NLSP, are the GMSB scenarios in which a neutral NLSP (usually a neutralino) radiatively decays into a gravitino LSP:
\begin{equation}
	\tilde\chi^0_1 \to \gamma\tilde G, \label{eq:delayed-GMSB}
\end{equation}
or a charged NLSP (usually the right-handed stau, $\tilde\tau_R$) decays into a $\tau$ and a gravitino NLSP:
\begin{equation}
	\tilde\tau_R \to \tau\tilde G \label{eq:delayed-GMSB-tau}
\end{equation}
In the GMSB scenarios the gravitino mass is given by~(see, e.g., \cite{Feng}):
$$
	M_{\tilde G} = \sqrt{\frac{8\pi}{3}}\frac{F}{M_P},
$$
where $F$ is the vacuum expectation vaue of the dynamical supersymmetry breaking, and $M_P$ is Planck mass. Since most of the GMSB models predict $F \sim 10^[14]$~GeV$^2$, and $M_P \sim 10^{19}$~GeV, the gravitino is expected to be very light. The fact that the gravitino interacts with matter only weakly, could make the decays (\ref{eq:delayed-GMSB}) and (\ref{eq:delayed-GMSB-tau}) very slow. For example, the $c\tau$ for the decay (\ref{eq:delayed-GMSB-tau}) is given by~\cite{Feng} :
$$
	c\tau \approx 10~\mbox{km} \times \langle\beta\gamma\rangle \times \left[\frac{\sqrt{F}}{10^7~\mbox{GeV}}\right]^4 \times \left[\frac{100~\mbox{GeV}}{m_{\tilde\tau_R}}\right]^5,
$$
and could be very large.

Generally speaking, in different SUSY models, the $c\tau$ of the charged or neutral NLSP can be from subatomic distances to many kilometers.

\section{Detection of the Delayed Decays}
\label{sec:delayed}

Detection and identification of long-lived particles depend on the decay path and the charge of this particle. A typical collider detector, e.g. the upgraded D\O\ apparatus~\cite{D0Upgrade}, has an inner Silicon Microstrip Tracker (SMT), capable of identifying secondary vertices from long-lived particle decays. The silicon detector is surrounded by a less precise Central Fiber Tracker (CFT), which is enclosed in the calorimeter. The calorimeter is surrounded by the muon system. Typical outer radii of the vertex detector, central tracking detector, calorimeter, and the muon system are $\sim 10$, 100, 200, and 1000~cm, respectively.

In the case of the charged long-lived particles, one can identify the secondary decay vertex in the silicon vertex detector for $\gamma c\tau$ between $\sim 0.1$~cm and $\sim 10$~cm. Since we expect SUSY particles to be heavy, in what follows we will assume $\gamma \approx 1$, i.e. $\gamma c \tau \approx c\tau$. Both the CDF and D\O\ experiments will be capable of identifying such secondary vertices and possibly trigger on them. For 10~cm $\lesssim c\tau \lesssim 100$~cm, the charged long-lived particle will predominantly decay inside the outer tracking volume. The resulting characteristic signature is a kink in the outer tracker and large $dE/dx$ in the silicon vertex tracker and the inner layers of the outer tracker, typical for a slow-moving charged particle. While triggering on the kinks in the outer tracker won't be possible in Run II, there is a good chance that such kinks can be found offline, if the event is accepted by one of the standard triggers. Since a massive slow moving particle still has a significant momentum, one would likely trigger on such events using a single high-$p_T$ track trigger, or a designated $dE/dx$ trigger. For 100~cm $\lesssim c\tau \lesssim 200$~cm, the charged long-lived particle will decay inside the calorimeter, giving a jet-like energy deposition inside it. Identification of these particles will therefore rely on the fact that such a jet has only one track pointing to it (similar to one-prong $\tau$-decays); moreover, this track will have high $dE/dx$. In the case of the D\O\ detector, an additional $dE/dx$ measurement in the preshower detector can be used to aid the identification of such particles both offline and at the trigger level. A single high-$p_T$ track trigger, a designated $dE/dx$ trigger, or a $\tau$-trigger  could be used to trigger on such events. Finally, for $c\tau \gtrsim 200$~cm, the long-lived particle will look stable from the point of view of the detector, and its identification will rely on a high $dE/dx$ track in the silicon vertex detector, the outer tracker, the calorimeter, and the muon system (if the $c\tau$ exceeds it outer radius). An additional time-of-flight (TOF) information from a designated TOF system (CDF) or muon system scintillators (D\O) could be used to trigger on such events and identify them offline. To summarize, both the CDF and D\O\ detectors will have very good capabilities for searches for charged long-lived particles with lifetimes from a few tens of picoseconds to infinity. Identification of the slow-moving particles using $dE/dx$ techniques in the case of the D\O\ detector is discussed in more detail in Section~\ref{sec:HIT-D0}.

The situation is much more complicated for a neutral weakly interacting particle. First of all, if such a particle decays outside of the calorimeter, it can not be detected, and will look like a missing $E_T$ in the event, i.e. like the LSP. It's unlikely that one could identify the presence of a long-lived NLSP in such events, but it still might be possible to identify these events as the non-SM ones. If events with such a signature are found in Run II, one could conceivably install a wall of scintillating detectors far away from the main detector volume and try to look for a photon from the radiative decay in this additional scintillator~\cite{Gunion}. For the purpose of this report, however, we will focus only on the case of $c\tau < 2$~m, which roughly coincides with the outer radius of the D\O\ calorimeter. The detection technique for such decays relies heavily on the fine longitudinal and transverse segmentation of the preshower detector and the calorimeter, which is an essential and unique feature of the D\O\ detector.

The signature for the radiative decay of a long-lived particle (e.g., (\ref{eq:delayed-GMSB}) or (\ref{eq:delayed-neutralino})), is a production of a photon with a non-zero impact parameter. If there was a way to identify the point of photon origin, one could single out such a delayed radiative decay corresponding to a very distinct and low background topology. As is shown in Section~\ref{sec:pointing}, the D\O\ calorimeter information, combined with the preshower information, can be used to achieve a very precise determination of the photon impact parameter. This technique would allow to identify long-lived neutral particles with 10~cm $\lesssim c\tau \lesssim 100$~cm, in the case of the upgraded D\O\ detector. The detectable $c\tau$ range can be further extended by a factor of two by looking for the photons from the radiative decay in the D\O\ hadronic calorimeter. The signature for such photons would look like a ``hot cell,'' i.e. an isolated energy deposit in one or two hadronic calorimeter cells. The reason that the EM energy deposition in the calorimeter is so isolated is the fact that each hadronic calorimeter cell contains many radiation lengths and completely absorbs an EM shower. The main background for this signature is production of high-$p_T$ $K_S^0$-mesons which decay into a pair of $\pi^0$'s corresponding to the 4$\gamma$ final state. Since the $K_S^0$ is significantly boosted, these four photons are highly collimated and will be identified as a single EM shower in the calorimeter. For a typical $K^0_S$ momentum of 25 GeV, which corresponds to $\sim 20$~GeV $E_T$ of the resulting $4\gamma$ system, the $\gamma c\tau$ is about 130~cm, i.e. most of the $K^0_S \to 4\gamma$ decays will occur in the hadronic calorimeter. This background, however, can be well predicted and also has a well defined $r$-dependence, where $r$ is the radial distance of the ``hot cell'' from the detector center. We therefore expect this background to be under control in Run II. Thus, the upgraded D\O\ detector will have unique capabilities for searching for neutral long-lived particles with 10~cm $\lesssim c\tau \lesssim 2$~m.

Both the CDF and D\O\ could also explore the 0.1~cm $\lesssim c\tau \lesssim 10$~cm range by looking for a conversion of the photon from a radiative decay in the silicon vertex detector. If the photon converts, one can determine its direction and the impact parameter fairly precisely by looking at the tracks from the $e^+ e^-$-pair. One, however, would pay a significant price for being able to explore this region of $c\tau$, since the probability of photon conversion in the silicon vertex detector is only a few per cent. CDF could in principle use the conversion technique to explore higher values of the $c\tau$ by looking for conversions in the outer tracking chamber. However, due to the low conversion probability the sensitivity of the CDF detector in this region is by far lower than that of the D\O\ detector.

Apart from being crucial for the study of physics with delayed radiative decays, the D\O\ detector ability for photon pointing is very attractive from other points of view. In the high luminosity collider environment the average number of interactions per crossing exceeds one. (It can be as high as five, for the 396~ns bunch spacing expected at the beginning of Run 2.) Therefore, each event will generally have several primary vertices with only one being from the high-$p_T$ interaction of physics interest, and the others being due to minimum bias $p\bar p$ collisions. The presence of multiple vertices creates a problem in choosing the right one for determination of the transverse energies of the objects. Photon pointing can solve this problem. It is especially important at the trigger level when the information about all the objects in the events is not generally available, and therefore high-$p_T$ objects which produce tracks in the tracking chambers can not always be used to pinpoint the hard scattering vertex. In some cases, for example the $\gamma + \MET$ final state, there are no objects with tracks at all, so there is no way to determine what vertex the photon originated from without utilizing the calorimeter-based pointing. Not knowing which vertex is the one from high-$p_T$ collision results in the object $E_T$ mismeasurement, which is especially problematic for missing transverse energy calculations. Indeed, \MET\ calculations rely heavily on the vertex position and picking the wrong one may result in significant missing transverse energy calculated in the event which in fact does not have any physics sources of real \MET. Not does only this affect physics analyses for topologies with \MET, especially the $\gamma + \MET$ one, but it also results in a worsening of the \MET\ resolution, hence a slower turn-on of the \MET-triggers and ergo higher trigger rates. It is, therefore, very important to have a way of telling the high-$p_T$ primary vertex, and photon pointing is the only way of doing this in the $\gamma + \MET$ case.

Subsequent sections contain technical detail on the high-$dE/dx$ and delayed photon identification.

\section{Photon Pointing at D\O} 
\label{sec:pointing}

\begin{figure}[thb]
\vspace*{0.1in}
\centerline{\protect\psfig{figure=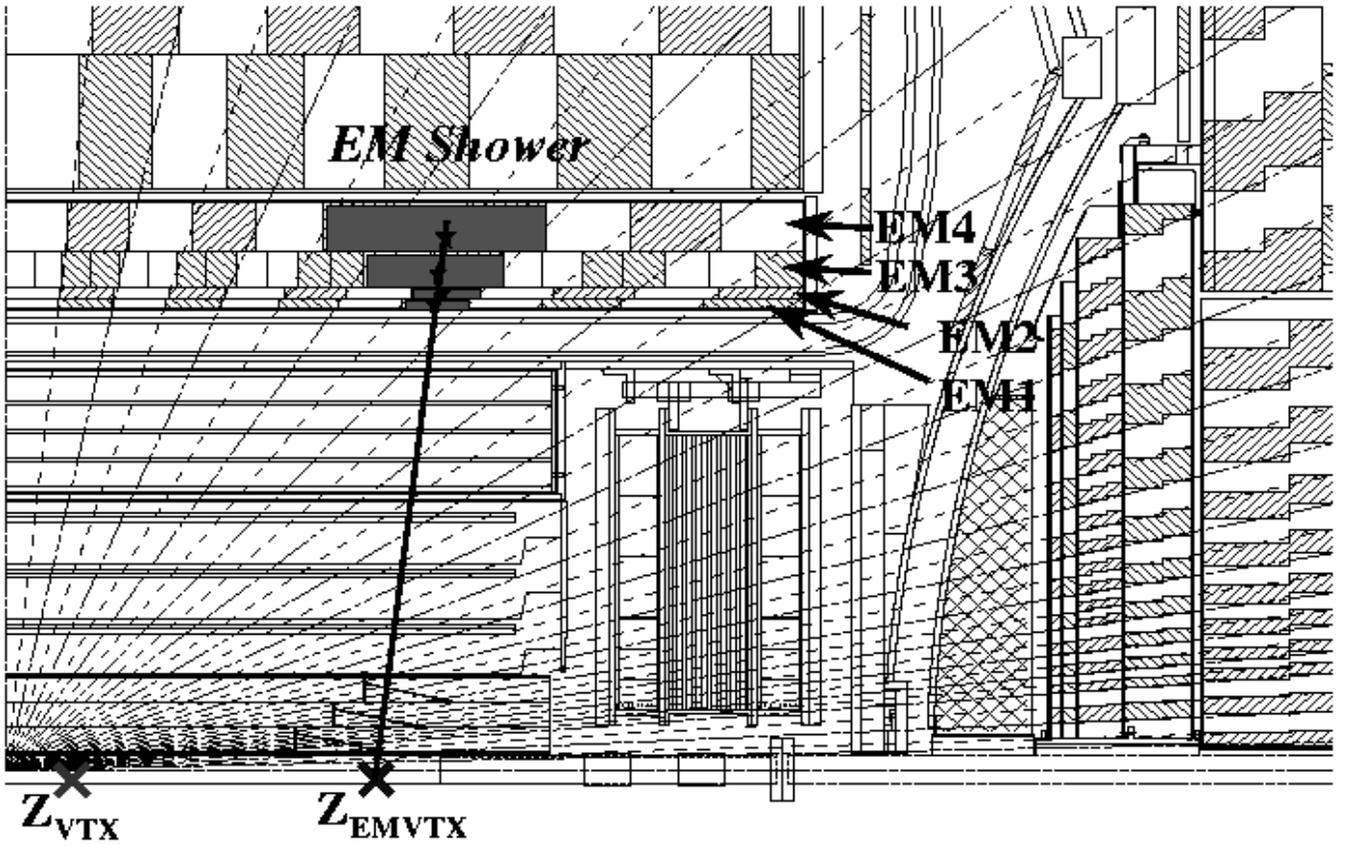,width=\textwidth}}
\caption{Side-view of the D\O\ calorimeter illustrating the application of the {\tt EMVTX} algorithm.}
\label{fig:emvtx}
\end{figure}

The importance of photon pointing was appreciated in some Run I analyses, particularly the studies of $Z(\nu\nu)\gamma$~\cite{Zg} and $\gamma\gamma$~\cite{monopoles}. We have utilized the fine longitudinal segmentation of the D\O\ EM calorimeter (which has four longitudinal layers) by calculating the c.o.g. of the shower in all four layers independently and then fitting the four spatial points to a straight line in order to determine the impact parameter and the $z$-position of the photon point-of-origin~\cite{EMVTX} (see Fig.~\ref{fig:emvtx}). The algorithm used for the c.o.g. finding is based on a logarithmic weighting of the energy deposition in the EM calorimeter cells which belong to the EM shower. The spatial resolution of the c.o.g. finding algorithm in four calorimeter layers averaged over central (CC) and forward (EC) rapidity range, as well as the geometrical parameters of the calorimeter layers are given in Table~\ref{table:EMVTX}. This study resulted in an algorithm, {\tt EMVTX}~\cite{EMVTX}, that has been used in several D\O\ analyses involving photons~\cite{Zg,monopoles}.

\begin{table}[htb!]
\begin{center}
\caption{Geometry and average resolutions in the preshower detectors and the EM calorimeter layers.}
\begin{tabular}{||l||l|l|l|l|l||}
\hline
Quantity 		& Preshower & EM1 & EM2 & EM3 & EM4 \\
\hline\hline
\multicolumn{6}{||c||}{Central Region}\\
\hline
$\langle R \rangle$ 
c.o.g. 		& 73.0~cm	& 85.5~cm	& 87.4~cm 	& 91.8~cm	& 99.6~cm	\\
$\sigma_z$ 		& 2.5~mm	& 20~mm	& 20~mm	& 7.0~mm	& 15~mm	\\
$\sigma_{r\phi}$ 	& 1.5~mm	& 17~mm	& 17~mm	& 3.5~mm	& 7.5~mm	\\
\hline
Quantity 		& CPS & EM1 & EM2 & EM3 & EM4 \\
\hline
\multicolumn{6}{||c||}{Forward Region}\\
\hline
$\langle |Z| \rangle$ 
c.o.g. 		& 142.0~cm	& 171.7~cm	& 174.2~cm	& 179.2~cm	& 189.7~cm  \\
$\sigma_r$ 		& 1.5~mm	& 8.0~mm	& 8.0~mm	& 1.5~mm	& 3.5~mm	\\
$\sigma_{r\phi}$ 	& 2.5~mm	& 7.0~mm	& 7.0~mm	& 1.0~mm	& 2.8~mm	\\
\end{tabular}
\end{center}
\label{table:EMVTX}
\end{table}

\begin{figure}[hbt]
\vspace*{-0.5in}
\centerline{\protect\psfig{figure=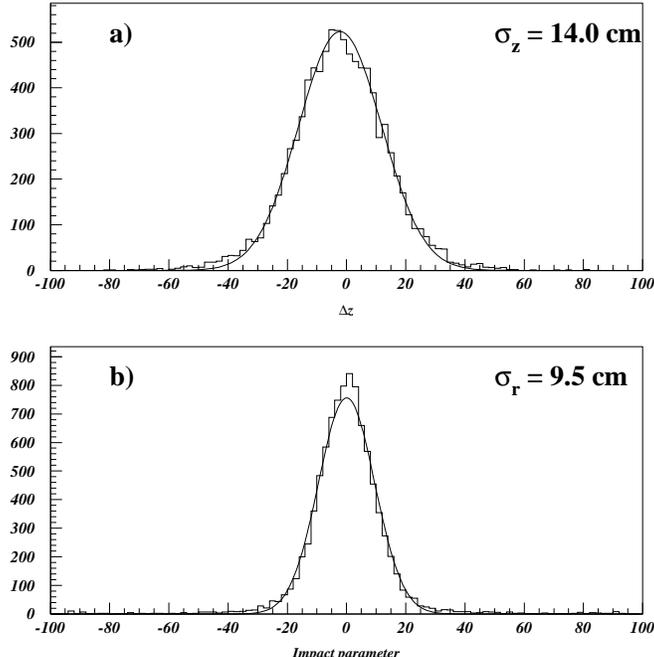,width=4.0in}}
\vspace*{-0.2in}
\caption{Comparison of (a) the error on the $z$-position of the vertex, and (b) the signed impact parameter from Run 1 $W \to e\nu$ data, with the results of toy simulations, for central electrons.}
\label{fig:emvtx_w}
\end{figure}

\begin{figure}[hbt]
\vspace*{-0.5in}
\centerline{\protect\psfig{figure=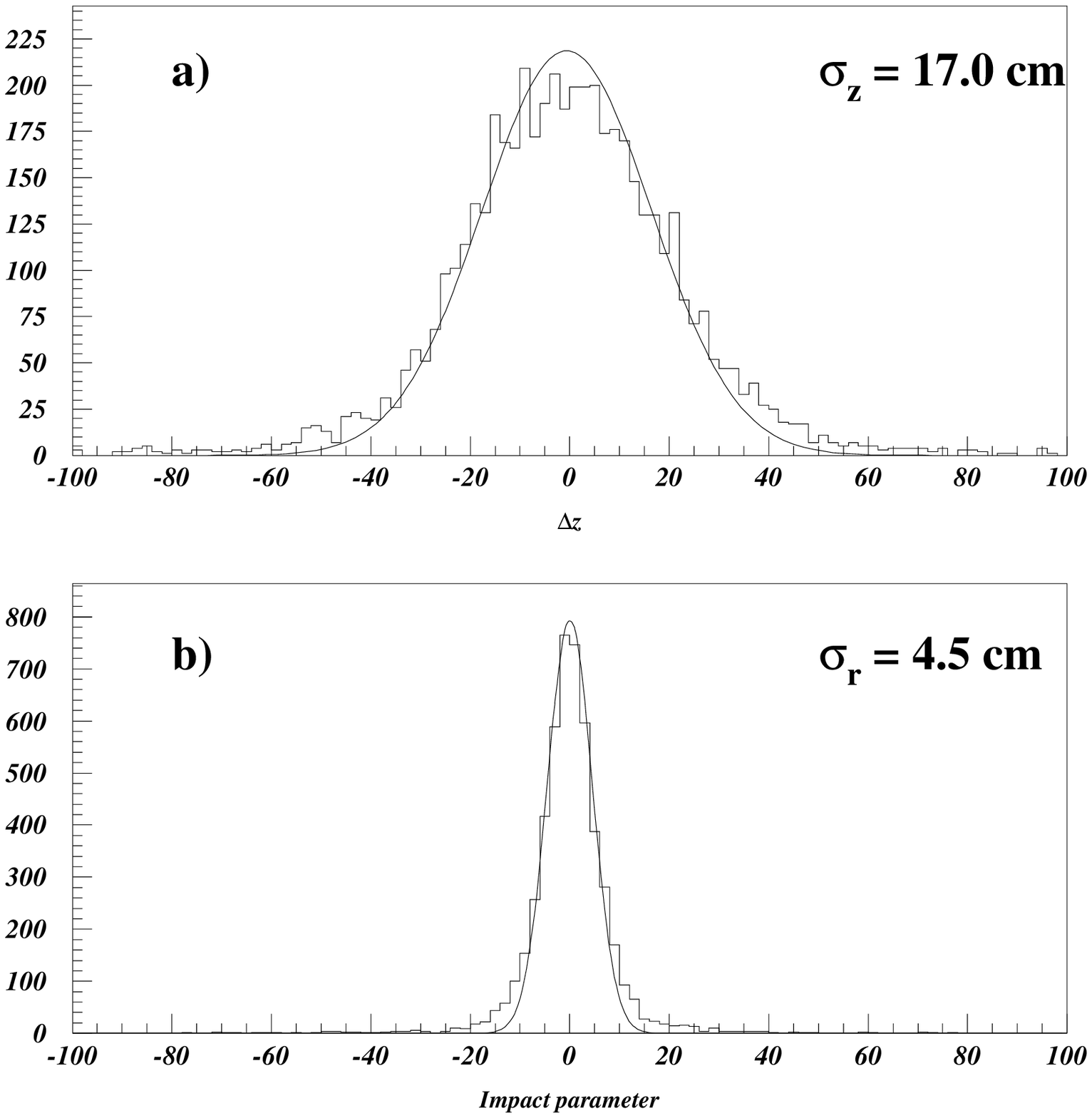,width=4.0in}}
\vspace*{-0.2in}
\caption{Comparison of (a) the error on the $z$-position of the vertex, and (b) the signed impact parameter from Run 1 $W \to e\nu$ data, with the results of toy simulations, for forward electrons.}
\label{fig:emvtx_w_ec}
\end{figure}

In order to study the improvement of photon pointing made possible by the utilization of the fine spatial resolution of the preshower detector, we have written a toy Monte Carlo (MC) simulation package which takes into account detector geometry, position error in calorimeter and preshower, as well as the primary vertex distribution. First, we compare the resolution obtained from the toy MC with the actual distributions obtained from $W \to e\nu$ events collected in Run 1. For electrons from $W$-events it is possible to determine the point of origin by using track information, which is quite precise. The difference between the $z$ position of the vertex obtained from tracking and from electron pointing as well as the signed impact parameter for the electron obtained by pointing are shown in Fig.~\ref{fig:emvtx_w}, for central electrons. (The positive sign corresponds to the impact parameter to the right of the center of the detector observed from the EM cluster location.) The distributions are fitted with Gaussian functions with the widths of $\sigma_z = 14.0$~cm and $\sigma_r = 9.5$~cm, as determined from toy MC. The data agrees well with the MC predictions and also appears Gaussian. An analogous comparison for forward electrons is shown in Fig.~\ref{fig:emvtx_w_ec}. The corresponding resolutions are $\sigma_z = 17.0$~cm and $\sigma_r = 4.5$~cm. The resolution on impact parameter improves in the forward region because the physical size of the calorimeter cells becomes smaller with an increase in $|\eta|$. In the $z$-direction, this effect is compensated by the small angle of the cluster pointing, which dilutes the $z$-resolution.


\begin{figure}[hbt]
\centerline{\protect\psfig{figure=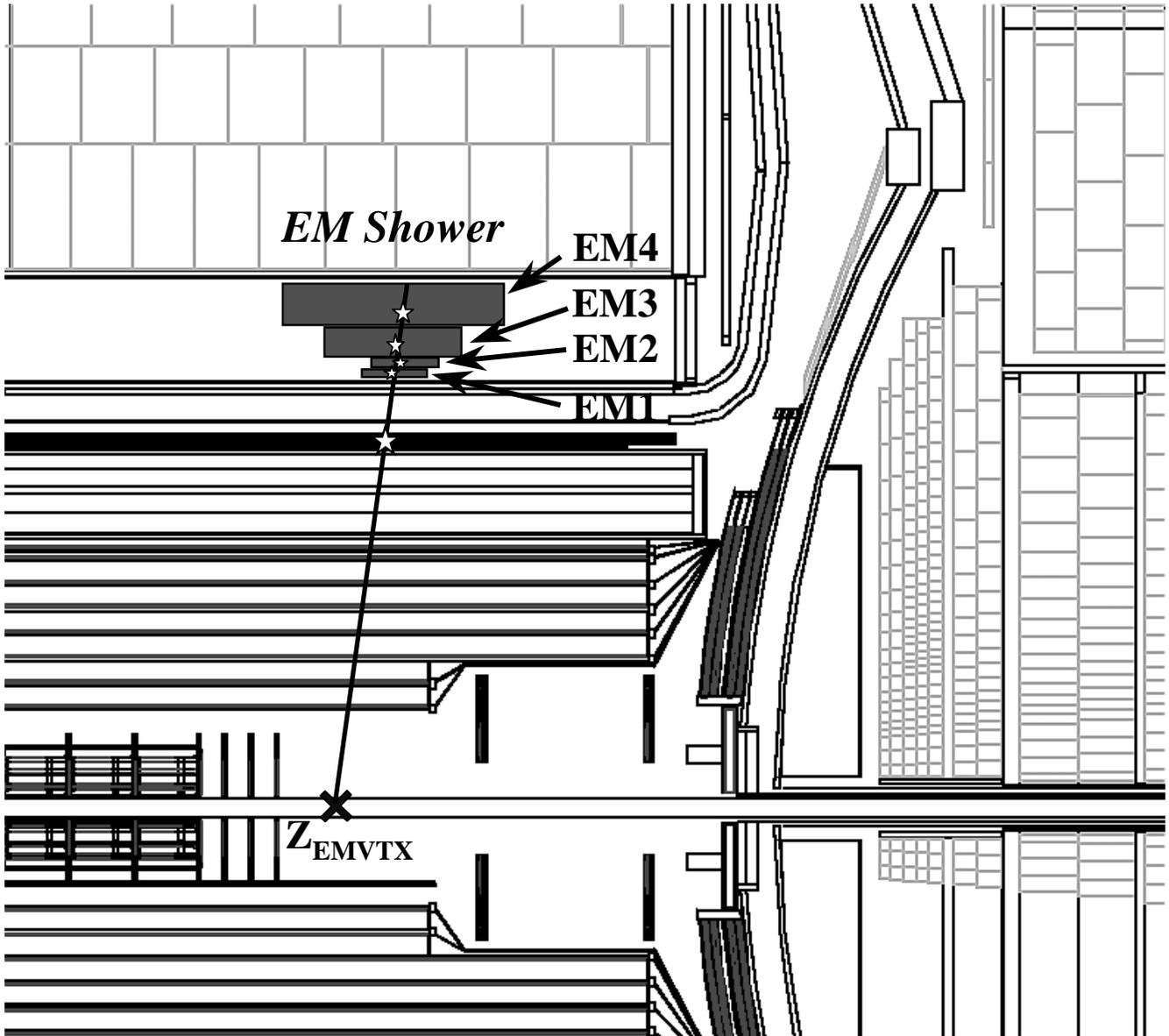,width=\textwidth}}
\caption{Side-view of the D\O\ calorimeter illustrating how the {\tt EMVTX} algorithm works.}
\label{fig:emvtx2}
\end{figure}
\begin{figure}[hbt]
\vspace*{-0.5in}
\centerline{\protect\psfig{figure=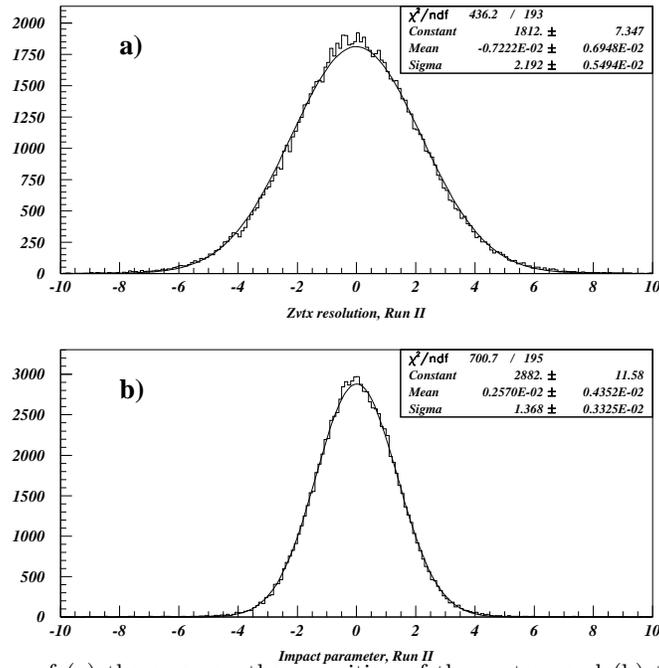,width=4.0in}}
\vspace*{-0.2in}
\caption{Run 2 simulation of (a) the error on the $z$-position of the vertex, and (b) the signed impact parameter, for central photons.}
\label{fig:emvtx_run2}
\end{figure}
\begin{figure}[hbt]
\vspace*{-0.5in}
\centerline{\protect\psfig{figure=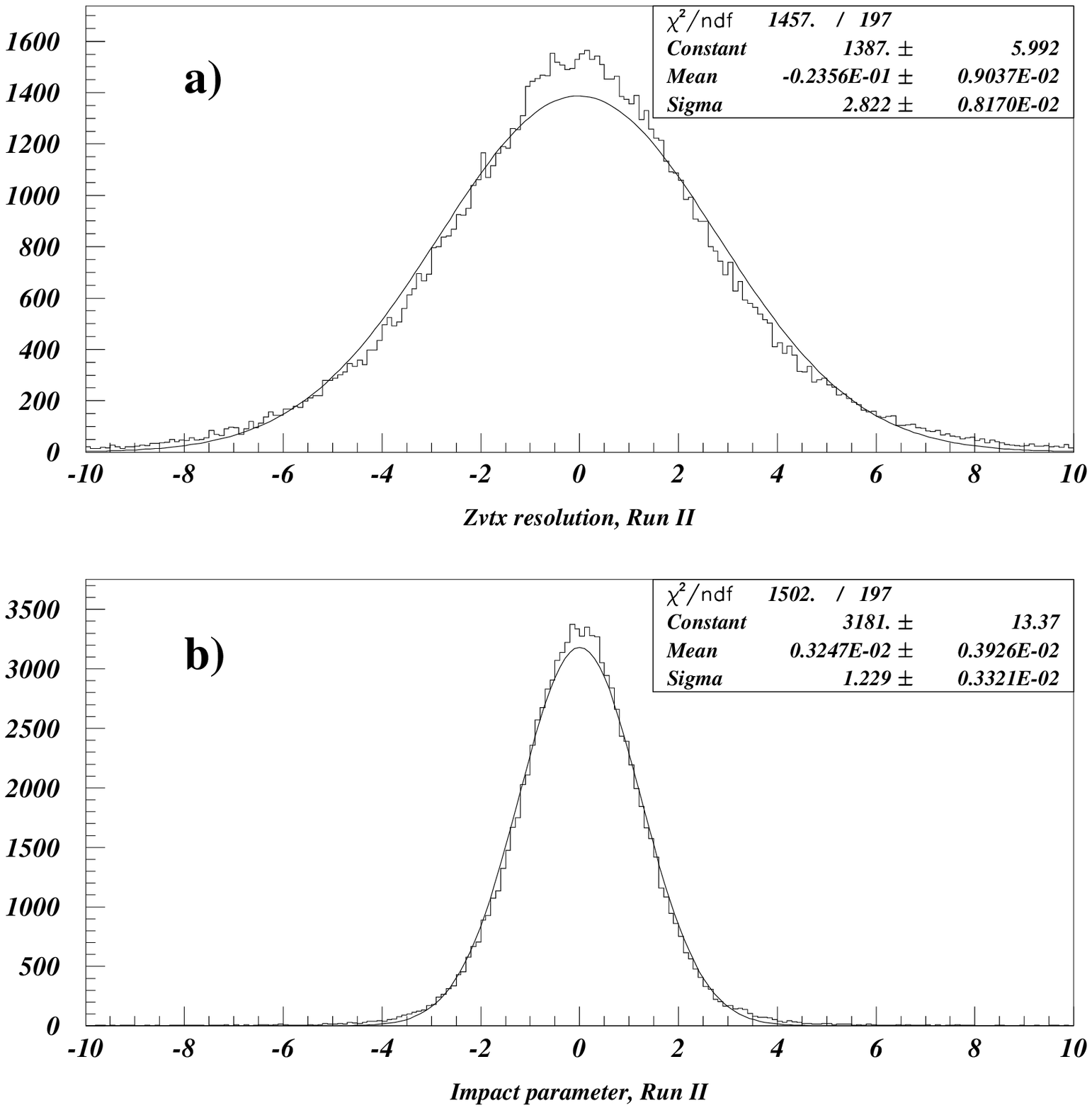,width=4.0in}}
\vspace*{-0.2in}
\caption{Run 2 simulation of (a) the error on the $z$-position of the vertex, and (b) the signed impact parameter, for forward photons. The slightly non-Gaussian shape is due to the change in the pointing resolution as a function of the photon rapidity.}
\label{fig:emvtx_run2_ec}
\end{figure}

The central and forward preshower detectors of the D\O\ Upgrade provide a precision measurement of the photon cluster position. The resolution for a typical EM shower perpendicular to the preshower strip is $\sim 1$~mm. Taking into account the crossing angle between the $u$- and $v$-planes in the preshower detectors, the following resolutions in $r\phi$ and $z$ ($r$) can be obtained: $\sigma_{r\phi} = 1.5$~mm, $\sigma_z = 2.5$~mm (CPS) and $\sigma_{r\phi} = 2.5$~mm, $\sigma_r = 2.5$~mm (CPS). As one can see, the preshower cluster position measurement is superior to that obtained from the EM calorimeter; its position is also spatially separated from that in the calorimeter, as seen in Fig.~\ref{fig:emvtx2}.

With the additional position measurement coming from the preshower detectors, the following pointing resolutions can be obtained for central and forward photons in Run 2: $\sigma_z = 2.2$~cm, $\sigma_r = 1.4$~cm (CC, see Fig.~\ref{fig:emvtx_run2}) and $\sigma_z = 2.8$~cm, $\sigma_r = 1.2$~cm (EC, see Fig.~\ref{fig:emvtx_run2_ec}). A very significant improvement in photon pointing (by a factor of six) is achieved by utilizing the additional position measurement provided by preshower detectors.

The implementation of the photon-pointing algorithm can be done as early as at the trigger Level 3. An approximate algorithm that uses only the c.o.g. of the EM shower in the preshower and in the third layer of the EM calorimeter could be used to decrease the amount of calculations and, hence, the decision-making time. Monte Carlo simulations show that the impact parameter and $z$-resolutions for a simplified algorithm are only 10\% worse than those obtained by complete five-point fit, which is quite satisfactory for trigger purposes. Having the precise vertex information from photon-pointing for photon triggers at Level 3, we will also recalculate the \MET based on this vertex, and that would significantly improve the turn-on of the \MET part of less inclusive triggers which would require an EM cluster and \MET.

\section{Detection of Slow-Moving Massive Charged Particles in D\O}

As described in Section~\ref{sec:delayed}, long-lived charged particles have a variety of characteristics which enable their identification, particularly in the analysis stage when the event's complete data set is available.  Depending on the lifetime, a combination of time-of-flight, ionization ($dE/dx$) in several different detectors, and the muon-like penetration of a high momentum, isolated track all can be employed, with the additional presence of a kink where there is a decay within the detector volume. Even though discovery at the analysis stage seems possible, triggering is really crucial, if heavy stable particles produced at the Tevatron are to be detected.  In this section we discuss the possibilities for triggering, and in particular, the use of $dE/dx$ in the hardware trigger as a tool for detecting these objects.

The D\O\ trigger system is hierachial, with 3 levels, each level passing a small subset of the events it examines to the next level for further analysis.  Thus Level 1 has a high input rate and examines a limited amount of information in making its decision, while subsequent levels have progressively lower input rates and spend longer analyzing more data.  At Level 3, all the data digitized for the event is available and the selection is made running software algorithms written in high level code and derived from the offline analysis.  We expect that the massive stable particles will be sufficiently rare, and their characteristics clear, such that separating candidates at Level 3 will be straight forward.  However, we need to understand how to indentify these events in the hardware triggers so that they survive to Level 3.

For Level 1 several tools are available to detect a massive stable particle.  The Central Fiber Tracker provides track candidates binned in momentum to which it can apply an isolation criteria~\cite{Yuri}.  The CFT has 80 trigger segments independently processed by the trigger. For a heavy stable particle one can select a high $p_T$ track with no other tracks in the home and adjacent segments, imposing an isolation of $\pm6.75^\circ$ in azimuthal angle~\cite{Yuri}.  Association of a muon with such a track would provide an efficient Level 1 trigger for massive stable particles, produced centrally and in isolation.

An additional tool~\cite{KJ} at Level 1 is the measurement of the time of flight (TOF) from scintillators associated with the muon detector and used primarily to reduce background in the large area muon detector from cosmic ray and beam associated accidentals.  These counters cover much of the area just outside the calorimeter, inside the first (``A") layer of muon detectors, and completely outside the detector, on top of the muon ``C" layer planes.  Electronics associated with the scintillation counters provides both a trigger gate and a TOF gate, thus giving time windows relative to the interaction time.  Scintillator hits received within the TOF gate will have the time of flight measurement digitized and saved with the data, assuming the event otherwise passes Level 1.  Typically the TOF gate will be set sufficiently wide to accept slow moving particles, namely at least 100 nsec.  To contribute to the Level 1 trigger, however, the scintillator hit must occur within the trigger gate, which is necessarily much shorter (of order 25 nsec) because of the high rate of accidentals, particularly in the ``A'' layer counters.  It may be possible to run with a considerable wider trigger gate for the ``C'' layer counters, given the lower expected accidental rate in these counters.  In this case, TOF from the muon system would provide a useful tool for Level 1 triggering on slow particles.  Given a Level 1 trigger, generated either through TOF or through an isolated stiff CFT track, the TOF data will be a useful tool for the Level 2 and Level 3 triggers.

\begin{figure}[hbt]
\vspace*{0.1in}
\centerline{\protect\psfig{figure=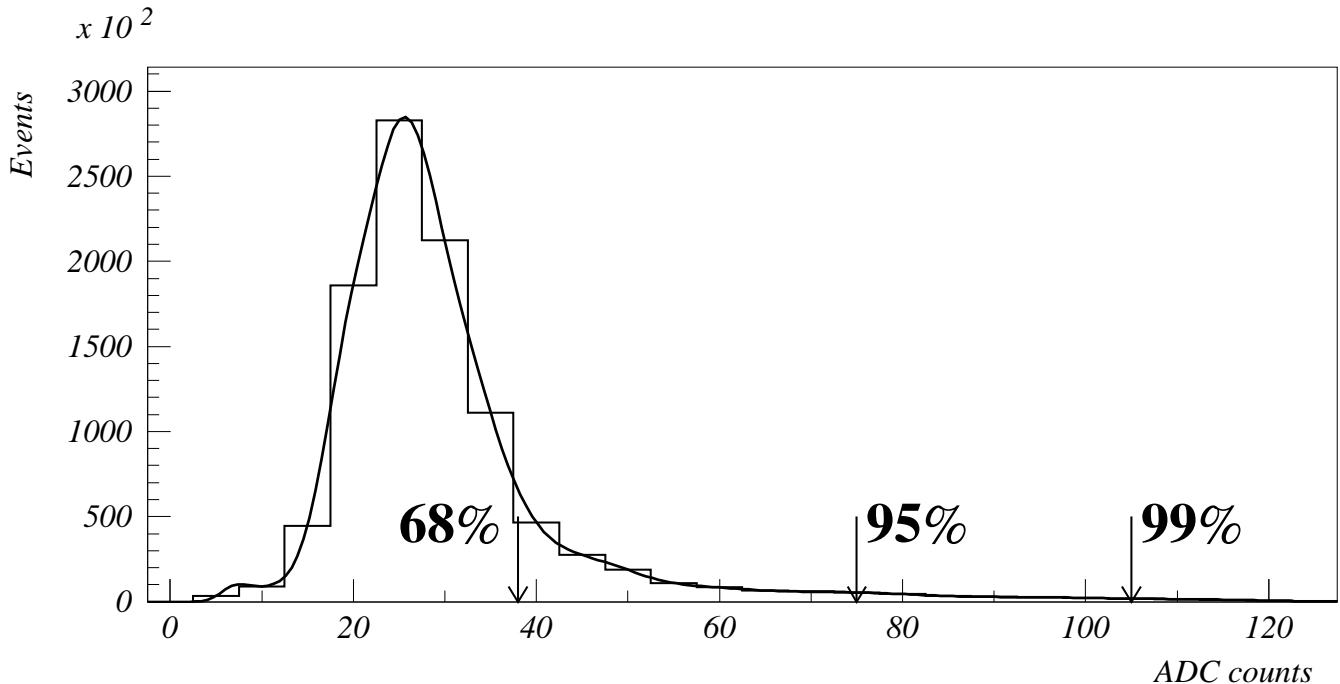,width=\textwidth}}
\caption{Energy deposition in one layer of the Silicon Microstrip Tracker in ADC counts, as measured in the test beam.}
\label{fig:MSP-ADC}
\end{figure}

Beyond Level 1, the new Silicon Microstrip Tracker provides interesting possibilities for triggering on slow moving particles.  Since energy loss of a slow moving particle drops as $1/\beta^2$ as a function of its velocity, the excellent $dE/dx$ energy resolution of the silicon chip provides a good handle for offline identification.  For example, the energy deposited in one silicon layer is shown in Fig.~\ref{fig:MSP-ADC}, as measured in a test beam~\cite{Maria}.    More importantly for the detection of slow moving particles, there are possibilities to exploit these measurements at the trigger level, with recent approval of the D\O\ Silicon Tracker Trigger (STT)~\cite{STT} as a component of the Level 2 trigger system.  The hardware design for the STT may allow for the inclusion of several additional backplane lines to carry $dE/dx$ information along with the other data associated with each cluster of hits~\cite{Uli}.  We have studied the basic capabilities of such trigger hardware to explore its potential for slow particle identification.

\begin{figure}[hbt]
\vspace*{0.1in}
\centerline{\protect\psfig{figure=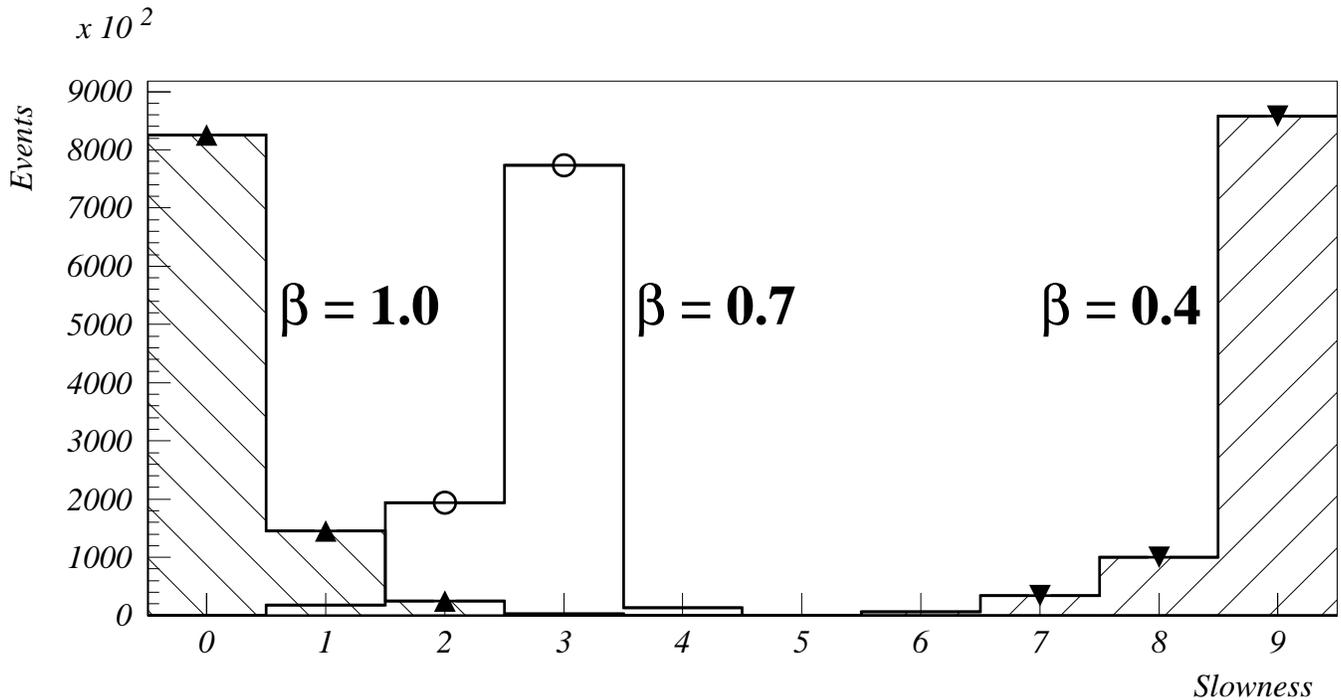,width=\textwidth}}
\caption{Distribution of slowness for massive particles with $\beta = 0.4$ and 0.7, and test beam particles with $\beta = 1.0.$}
\label{fig:MSP-slowness}
\end{figure}

Our simulation of the STT slow particle trigger is based on a simple Monte Carlo generator which returns an ADC value appropriate for the spectrum shown in Fig.~\ref{fig:MSP-ADC}.  We assume that two backplane lines per SMT hit are available, and we encode each ADC value into these two bits of data, or four bins.  After trying various values for the three bin edges, we have chosen to define the bins by those ADC values below which lie 67\%, 95\%, and 99\% of the data. These cutoffs correspond to 38, 75, and 105 ADC counts (see Fig.~\ref{fig:MSP-ADC}). The STT Level 2 trigger processor finds tracks using four layers of silicon; so, in our model, the hardware would provide four samples of this two bit $dE/dx$ data for each track.  At this point the trigger would use some algorithm to combine the four samplings most advantageously.  The major concern for the correct identification of a slow moving particle is the likelihood of false signals from a minimum ionizing particle (MIP), due to an occasional response in the very long tail of the Landau energy loss distribution.  Based on a few studies, our preliminary suggestion is simply to sum the three lowest values, rejecting the largest ADC count of the four.  The data then would provide a parameter, related to a $dE/dx$ of the particle in the silicon tracker, which we call ``slowness," and which has 10 possible values (0..9).  From our simulation we derive the distribution in slowness for a $\beta=1$ particle (equivalent to the test beam pion), as shown in Fig.~\ref{fig:MSP-slowness}.  

The ADC response to the passage of a slow moving particle will be similar to the ADC distribution of Fig.~\ref{fig:MSP-ADC} scaled by the factor $1/\beta^2$.  However, because of the very different kinematics (the slow moving particle is massive, typically 150 GeV) the energy loss distribution will have a less pronounced Landau tail.  We make a very conservative assumption and use only the Gaussian component in generating ADC values for massive slow moving particles.  Several resulting distributions in our $dE/dx$ trigger parameter, ``slowness,'' for particles with $\beta=0.4$ and $\beta=0.7$, are included in Fig.~\ref{fig:MSP-slowness}. There is a clear separation in this parameter compared to the $\beta=1$ distribution.

\begin{figure}[hbt]
\vspace*{0.1in}
\centerline{\protect\psfig{figure=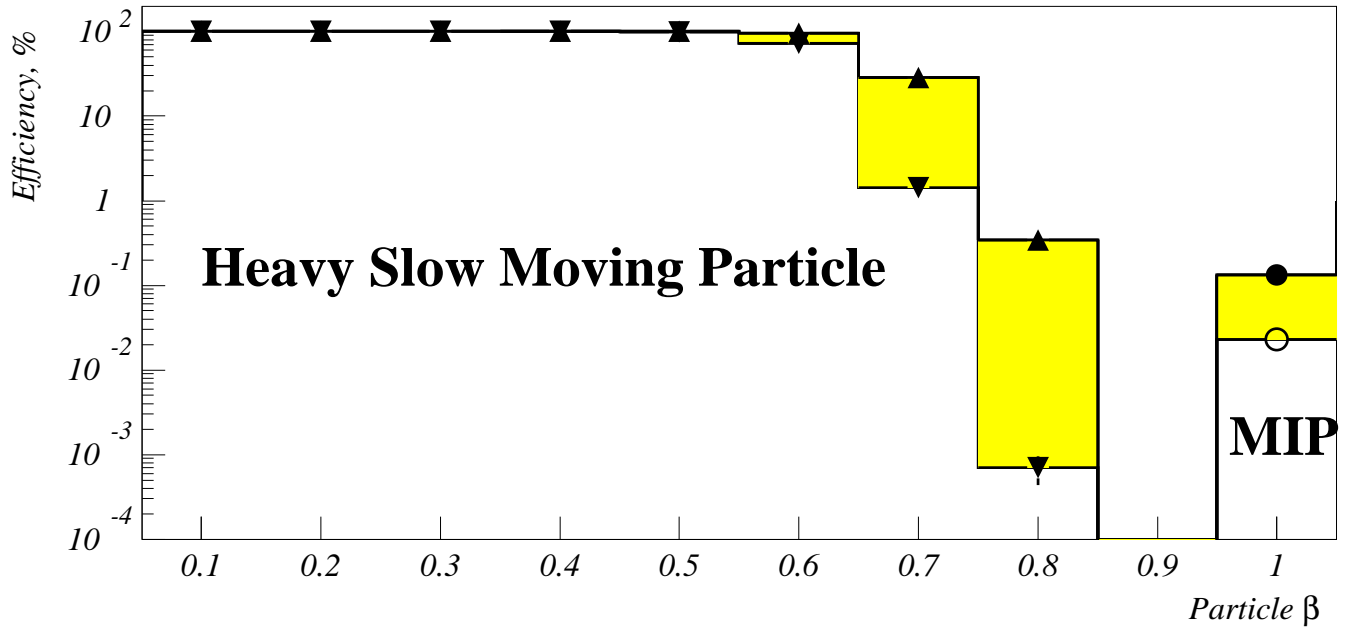,width=\textwidth}}
\caption{Efficiency as a function of particle $\beta$ for events with slowness $>3$. The open histogram corresponds to the case without angular smearing; the filled area shows the change when angular smearing of $\pm 45^\circ$ is taken into account.}
\label{fig:MSP-eff}
\end{figure}

To estimate the effectiveness of the STT $dE/dx$ trigger we consider a selection which tags as a slow particle those whose ``slowness" is greater than or equal to some value.  We vary this selection to study the efficiency for slow particles (at highest possible $\beta$) while maintaining a strong rejection against $\beta=1$ particles.  Figure~\ref{fig:MSP-eff} shows the acceptance as a function of a particle's velocity, for events with slowness $>4$.  The appearance of a few $\beta=1$ particles but not other high $\beta$ events reflects the conservative use, for massive particles, of a Gaussian ADC response, rather than the Landau distribution, which is used to generate the ADC values from a MIP.

\begin{figure}[hbt]
\vspace*{0.1in}
\centerline{\protect\psfig{figure=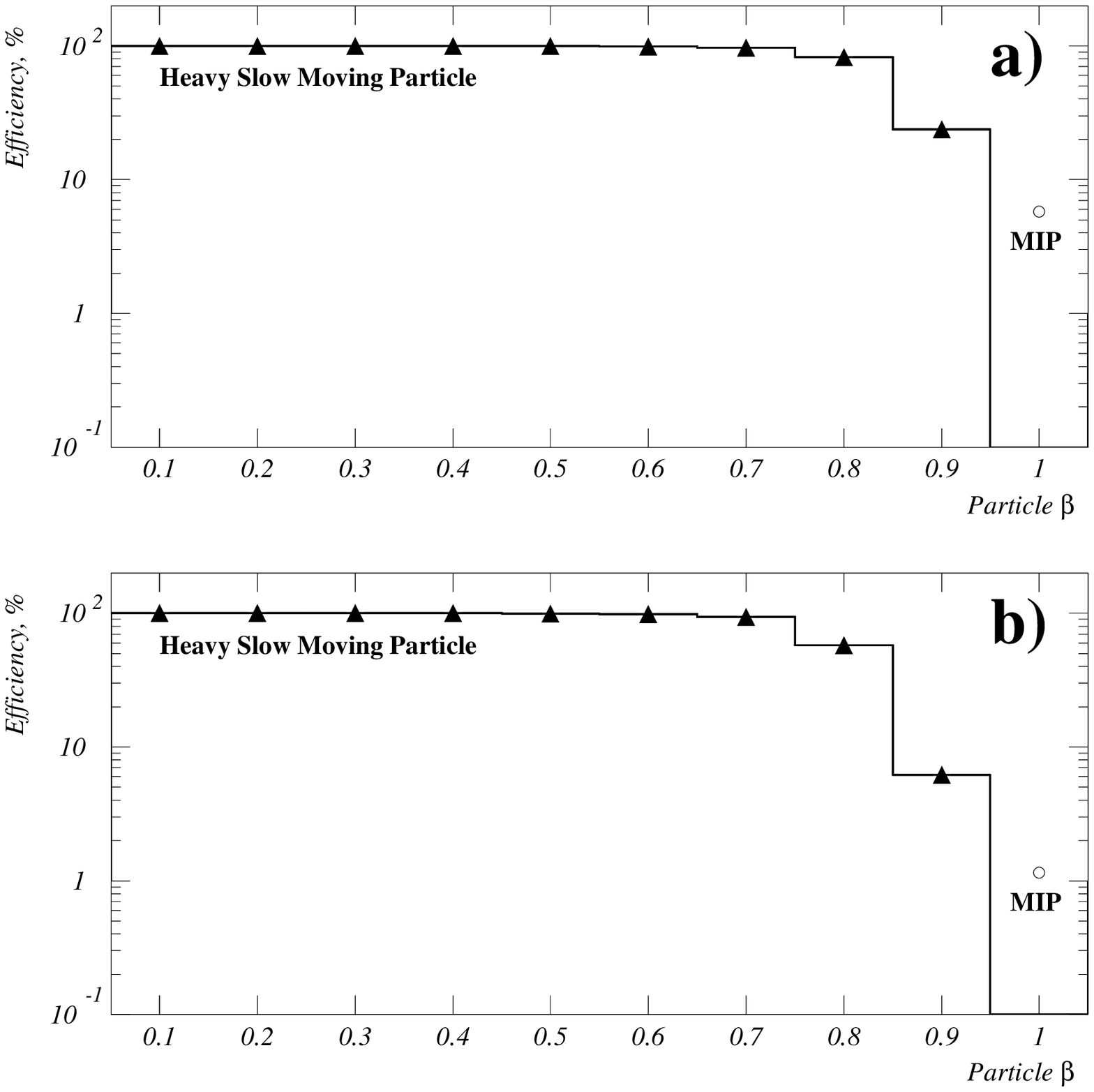,width=\textwidth}}
\caption{Efficiency as a function of particle $\beta$ for the events with redefined slowness a) $>7$; b) $>8$. Angular smearing of $\pm 45^\circ$ is taken into account.}
\label{fig:MSP-eff1}
\end{figure}

Despite the intent to trigger only on centrally produced objects, tracks will tend to be inclined to the silicon, increasing and broadening the $dE/dx$ response.  In fact, the STT hardware will search for track candidates from adjacent barrels; so that tracks may be inclined to the normal by as much as $45^\circ$ although the STT will have no information about this angle.  We have modeled the effect of inclined tracks by scaling the appropriate ADC response by $1/\cos\theta$, where $\cos\theta$ is generated uniformly between .707 and 1.0, or by $\eta$, where $\eta$ is generated uniformly between 0 and 0.88.  The differences between smearing based on $\cos\theta$ and that based on $\eta$ are small.  We present here results using $\cos\theta$ smearing for the centrally produced massive stable particles and $\eta$ smearing for the $\beta=1$ background.  As seen in Fig.~\ref{fig:MSP-eff}, the effect of the variation in ADC response due to track inclination is only a small reduction in the rejection for $\beta=1$ particles; moreover, this effect helpfully raises the cutoff in $\beta$ for massive particles.  Overall, this smearing does not appear to affect the STT ``slowness" trigger significantly. Including track inclination, the study suggests that the STT Level 2 $dE/dx$ trigger would provide a fully efficient tag for slow particles up to $\beta=0.7$ with an acceptance of $\beta=1$ particles less than $2\times 10^{-3}$.  

Because of its excellent rejection for $\beta=1$ particles, the Level 2 $dE/dx$ trigger will be a good means to select slow moving particles, independent of other criteria such as TOF or track isolation.  However, if the particle is sufficiently long-lived to traverse the entire detector, it may be possible to relax the $dE/dx$ requirement.  We have studied the effect of modifying the hardware ADC sampling, to explore widening the acceptance of heavy particles in $\beta$ at the expense of
$\beta=1$ rejection.  There is good physics motivation in doing so, as some GMSB models predict the production of heavy stable particles with $\beta$ in the range 0.8-0.9~\cite{Jianming}. Using ADC bins with edges corresponding to 50\%, 68\%, and 90\% of the distribution, we find good acceptance for massive particles at high $\beta$, as shown in Fig.~\ref{fig:MSP-eff1}, with rejection factors between 20 and 100, for $\beta=1$ particles.

The above study illustrates the potential of a STT $dE/dx$ trigger.  It does assume that the ADC distribution for a MIP is as seen in the test beam, and that differences in the silicon chip response over the detector won't significantly affect this distribution.  The study suggests that track inclination may not be a serious problem.  Futher, the ``slowness" flag derived with the STT from $dE/dx$ can in Level 2 be combined with other information (as of a straight, non-oblique and isolated muon) to provide a global Level 2 trigger.  In summary, it seems promising that the STT could provide very useful $dE/dx$ information early in a slow particle's lifetime, which can be combined with other data to create an efficient trigger for these interesting objects.

\label{sec:HIT-D0}

\end{document}